\documentclass[sigconf]{acmart}
\usepackage{wrapfig}
\usepackage{comment}
\usepackage{colortbl}

\usepackage{hyperref}

\usepackage{url}

\usepackage{amsthm} 

\usepackage{xcolor}

\definecolor{mycolor}{HTML}{66CCFF} 
\usepackage{color,soul}

\usepackage{graphics}
\usepackage{hyperref}
\usepackage{lipsum}
\usepackage{tikz}
\usepackage{enumitem}	
\usepackage{listings} 
\usepackage{booktabs}	
\usepackage{lipsum}

\usepackage{capt-of}	
\usepackage{diagbox}
\usepackage{multirow}
\usepackage{todonotes}  
\definecolor{refkey}{rgb}{249,158,26}
\definecolor{labelkey}{rgb}{0,1,0}
\definecolor{airforceblue}{rgb}{0.36, 0.54, 0.66}
\definecolor{applegreen}{rgb}{0.55, 0.71, 0.0}

\usepackage{mathrsfs}	
\usepackage{amsfonts}

\usepackage[export]{adjustbox} 

\usepackage[plain]{algorithm2e}
\usepackage{boldline}					

\usepackage{multirow}

\definecolor{frenzyorange}{RGB}{249, 158, 26}

\renewcommand{\paragraph}[1]{\vskip 3pt\noindent\textbf{#1 }}	 

  {\begin{list}{$\bullet$}%
     {\setlength{\parsep}{0pt}%
      \setlength{\topsep}{0pt}%
      \setlength{\itemsep}{2pt}}}%
  {\end{list}}
\newcommand\Noted[1]{} 

\definecolor{darkblue}{rgb}{0.0, 0.0, 0.55}
\definecolor{mygreen}{HTML}{ADFF2F}
\definecolor{mylightgray}{gray}{0.8}

\newenvironment{myitemize}%
  {\begin{itemize}
	[leftmargin=0cm,
		itemindent=.3cm,
		labelwidth=\itemindent,
		labelsep=0pt,
		parsep=1pt,
		topsep=1pt,
		itemsep=1pt,
		align=left]
  }%
  {\end{itemize}}    

  {\begin{enumerate}
	[leftmargin=.cm,itemindent=.5cm,labelwidth=\itemindent,
		labelsep=0pt,
		parsep=1pt,
		topsep=1pt,
		itemsep=3pt,
		align=left]
  }%
  {\end{enumerate}}    

\newcommand\sect[1]{Section~\ref{sec:#1}}	

\newcommand{\code}[1]{\texttt{\small{#1}}}	

\newcommand{\sys}{\code{WhisperFlow}}
\newcommand{\baseline}{\textit{Whisper-S}}
\newcommand{\stransformer}{\textit{S-Transformer}}
\newcommand{\ourss}{\textit{Ours}}
\newcommand{\oursp}{\textit{Ours-pipeline}}

\makeatletter
\def\@copyrightspace{\relax}
\makeatother

\copyrightyear{2025}
\acmYear{2025}
\setcopyright{cc}
\setcctype{by}
\acmConference[MobiSys '25]{The 23rd Annual International Conference on Mobile Systems, Applications and Services}{June 23--27, 2025}{Anaheim, CA, USA}
\acmBooktitle{The 23rd Annual International Conference on Mobile Systems, Applications and Services (MobiSys '25), June 23--27, 2025, Anaheim, CA, USA}
\acmDOI{10.1145/3711875.3729151}
\acmISBN{979-8-4007-1453-5/2025/06}

\begin{document}

\title{WhisperFlow: speech foundation models in real time}



\author{Rongxiang Wang}
\affiliation{%
	\institution{University of Virginia}
	\city{}
	\state{}
	\country{}
}
\email{waq9hw@virginia.edu}

\author{Zhiming Xu}
\affiliation{%
	\institution{}
	\city{}
	\state{}
	\country{}
}
\email{zhimng.xu@gmail.com}

\author{Felix Xiaozhu Lin}
\affiliation{%
	\institution{University of Virginia}
	\city{}
	\state{}
	\country{}
}
\email{felixlin@virginia.edu}


\date{}

\thispagestyle{empty}

\begin{abstract}

Speech foundation models, such as OpenAI's Whisper, become 
the state of the art in speech understanding due to their strong accuracy and generalizability. 
Yet, their applications are mostly limited to processing pre-recorded speech, 
whereas processing of \textit{streaming} speech, in particular doing it efficiently, remains rudimentary. 
Behind this inefficiency are multiple fundamental reasons: 
(1) speech foundation models are trained to process long, fixed-length voice inputs (often 30 seconds); 
(2) encoding each voice input requires encoding as many as 1,500 tokens with tens of transformer layers; 
(3) decoding each output entails an irregular, complex beam search. 
As such, streaming speech processing on resource-constrained client devices is more expensive than other AI tasks, e.g., text generation.

To this end, we present a novel framework, \sys{}, which embodies both model and system optimizations. 
(1) \textit{Hush word} as a short, learnable audio segment;  appended to a voice input, a hush word gracefully stops the speech model from processing more input without  hallucination; 
(2) \textit{Beam pruning}, which aligns streaming audio buffers over time and reuses results from earlier decoding rounds, therefore significantly accelerating decoding; 
and (3) \textit{CPU/GPU pipelining}, which not only maps to the encoding/decoding stages dynamically, but also tunes to an optimal resource ratio,  
respecting the encoding/decoding speed that varies across voice inputs, models, and hardware.

We test \sys{} on commodity ARM platforms with 4--12 CPU cores and 10–-30 GPU cores.
It reduces per-word latency by 1.6$\times$--4.7$\times$ to as low as 0.5 second, 
while seeing negligible accuracy degradation.
On an entry-level MacBook Air, 
\sys{} can keep the per-word latency around 1 second, with the whole device drawing only 7 Watts in total.
\end{abstract}

\ccsdesc[500]{Information systems~Mobile information processing systems}
\ccsdesc[500]{Computing methodologies~Machine learning}

\maketitle
\pagestyle{plain}

\section{Introduction}
\label{sec:intro}



\paragraph{Streaming speech processing (SSP)}
This paper addresses automatic speech recognition on streaming inputs: 
as users speak, a computer system continuously transcribes the speech into uttered words in real-time~\cite{he2019streaming, moritz2020streaming, kannan2019large}. 
Such a task, known as streaming speech processing (SSP),
is at the core of various applications including transcribing daily conversations~\cite{xiong2017toward}, live-stream transcriptions~\cite{ranchal2013livestream}, online meetings~\cite{tur2008calo}, medical documentation~\cite{blackley2019speech}, and note-taking~\cite{shadiev2014review}. 
Both accuracy (e.g., medical transcription) and latency (e.g., live meeting captioning) are critical for these applications.
Providing a good accuracy-latency trade-off is crucial for making SSP practical on client devices.

\begin{figure}[t!]
	\centering
	\includegraphics[width=0.48\textwidth]{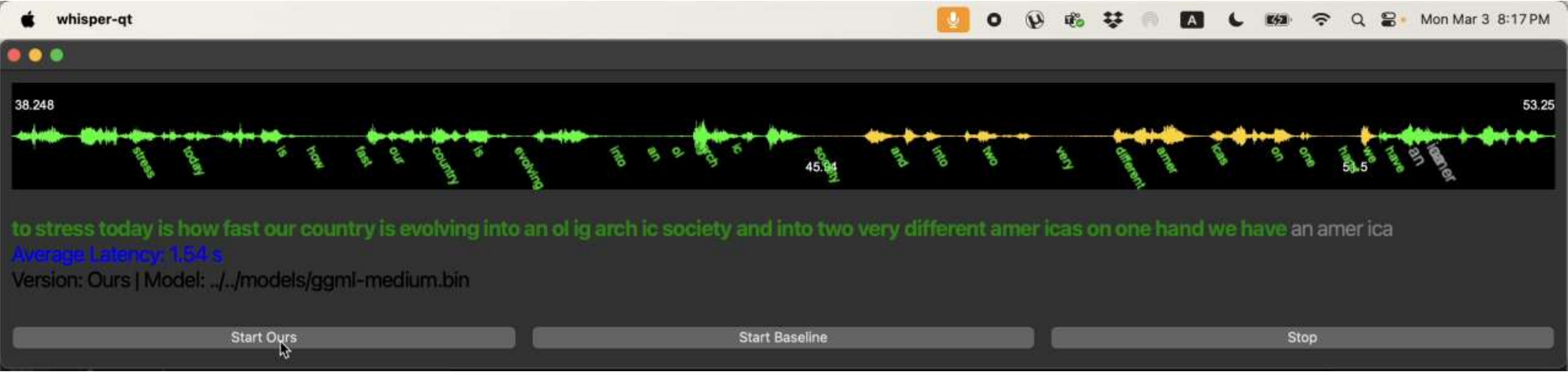}
	\caption{\sys{} running on Apple M2, which is evaluated in \sect{eval}
    }
	\label{fig:demo}	
\end{figure}


\begin{figure}[t!]
	\centering
	\includegraphics[width=0.48\textwidth]{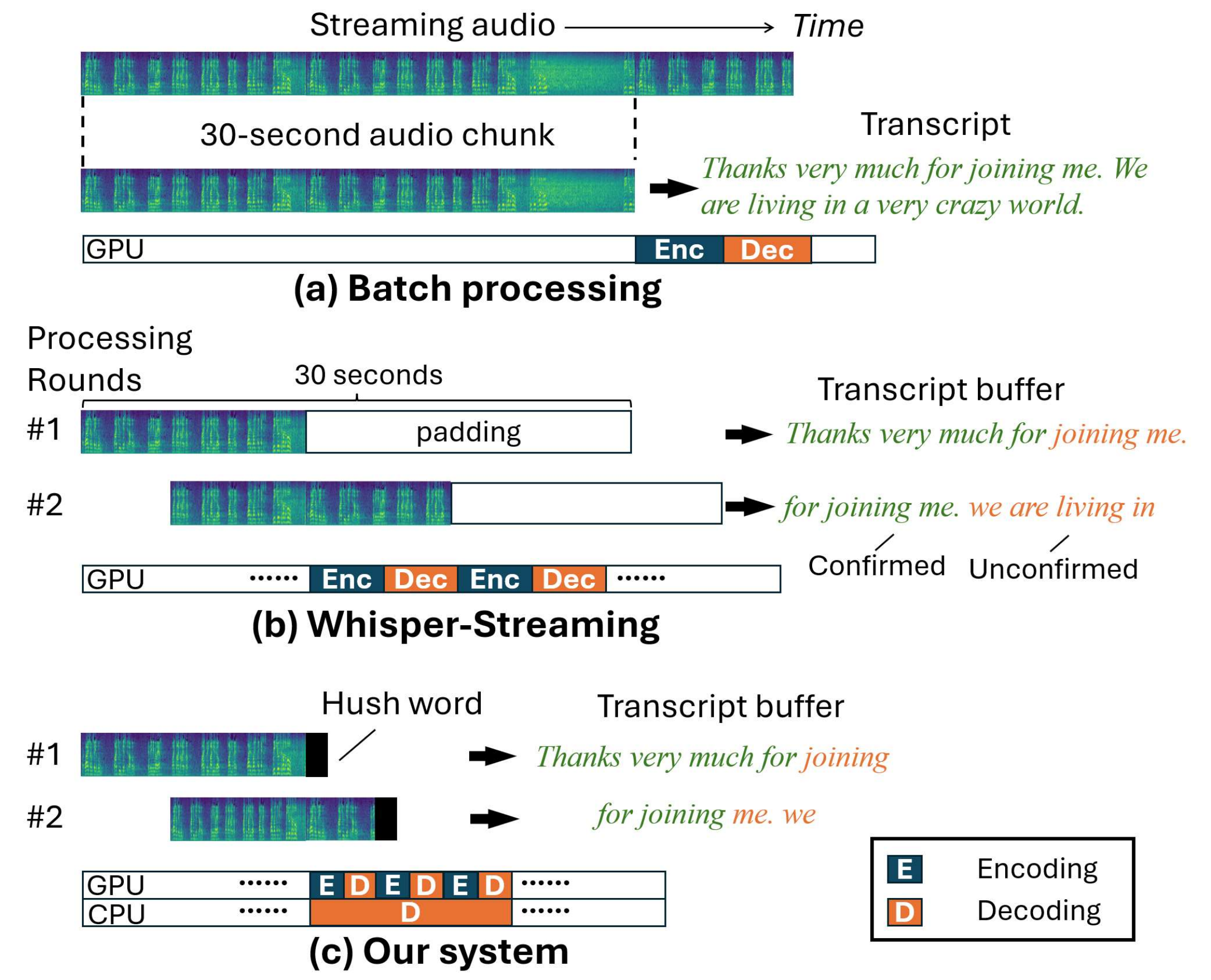}
 	\caption{A comparison of approaches~\cite{radford2023robust,machavcek2023turning} for running a speech foundation model over streaming input 
    }
	\label{fig:intro}
	
\end{figure}
\subsection{Why SSP is challenging}
 Speech understanding has been revolutionized by foundation models. 
 Notably, OpenAI publishes Whisper ~\cite{radford2023robust},  a family of encoder-decoder \textit{transformer} models. 
 Unlike traditional speech models such as \textit{Conformer} models~\cite{gulati2020conformer}, 
 the foundation models are pre-trained on a much larger dataset (680K hours of audio data), 
 directly map audio to text, and generalize better on unseen speech inputs.

Although speech foundation models already sparkle extensive research on accuracy and efficient execution, 
little has been done on executing them efficiently on streaming inputs, for reasons below: 

First, foundation models are ingrained to handle fixed-length inputs. 
For example, Whisper requires any input shorter than 30 seconds to be padded to 30 seconds before being fed to the model. 
Otherwise, the model's output may include hallucination 
(see Section~\ref{sec:motiv}).
This behavior is likely a deliberate decision to enable efficient pretraining on many GPUs, scaling up to colossal data (680K hours of audio for Whisper). 
A ramification, however, is that even encoding a short input, e.g., of 1 second, would require the same amount of computation as encoding a 30-second input.

Second, the model must encode a long context for each utterance: 
a 30-second voice input is discretized into 1500 tokens, 
which then pass through tens of transformer blocks.
This is shown in Figure~\ref{fig:intro} (a). 
For example, Whisper (medium, with 12 encoder layers) would require similar GFLOPS 
as a middle-sized GPT-2 model~\cite{radford2019language} processing a long text input with over 1000 words. 
Such a huge cost is unprecedented in GenAI on client devices.


Third, expensive, irregular decoding due to beam search algorithm~\cite{hayes1976speech}.
To overcome audio noises, 
the decoding process maintains multiple (often 5) auto-regressive decoding paths, each updating a tentative output sequence, called a hypothesis. 
A significant fraction of the beam search -- token scoring, ranking, and hypothesis pruning, has irregular parallelism; they are \textit{unamenable} to GPU execution. 
This is more complex than LLM generation which often uses a single path decoding. 

\paragraph{Existing approaches to SSP}
Existing approaches to SSP follow a common paradigm: evaluating a foundation model repeatedly on audio inputs as they arrive. 
Whisper.cpp~\cite{ggerganov2022whispercpp}, an efficiency-oriented implementation of Whisper, supports a simplistic streaming mode that processes audio in small, disjoint segments (each padded to 30 seconds). 
It is unsuitable for deployment due to significant word errors.
Recent research, Whisper-Streaming~\cite{machavcek2023turning}, processes incoming audio in \textit{sliding windows}, where each window comprises audio samples padded to the model's fixed input length. The execution pipeline only materializes the output tokens after they are confirmed by the model's Local-Agreement2 mechanism~\cite{Liu2020LowLatencySS}, resulting in fewer word errors than the naive streaming mode. 
In exchange, the execution now demands much higher GFLOPS. As shown in ~\autoref{fig:intro} (b), the same audio samples are included in multiple consecutive windows, repeatedly participating in encoding and decoding computations.
However, eliminating such redundancy is non-trivial: 
one cannot simply compose earlier processing results to get a later window's results, due to complex challenges, which will be detailed in ~\autoref{sec:motiv} and ~\autoref{sec:design1}. As a result, existing approaches fail to strike a good balance between accuracy and latency, making them impractical for real-world scenarios. For example, live-stream transcription of critical news events demands both high fidelity and low latency; current systems force a trade-off where low latency risks higher error rates—and potential misinformation—while high accuracy results in delays that leave audiences unable to keep up with the stream.

\subsection{Our key ideas}
This paper focuses on accelerating SSP, making it suitable for resource-constrained devices. 
We achieve this through a combination of model optimization and execution optimization:
the former reduces the total GFLOPS required by the model, 
whereas the latter maps the resultant model inference stages to modern hardware. 


Our first idea is to eschew the input padding by appending a short audio segment (dubbed ``hush word'') to the actual audio input.
The idea originates in security research~\cite{chen2020devil, raina2024mutingwhisperuniversalacoustic,raina2024controllingwhisperuniversalacoustic}, 
which injects a short input audio segment and hence prevents it from producing any token (i.e., ``mute'' the model).
We give this approach a novel, positive spin: 
instead of adversarial model manipulation, 
we gracefully stop the model from producing more tokens (i.e. ``hush'' it) 
and generate \textit{proper} outputs; 
unlike the attacks which \textit{prepend} the audio segment to the input,
we address challenges and demonstrate that the hush word can be \textit{appended} to an input without causing model hallucinations.
We show that such a hush word can be as short as 0.5 second (equivalent to only 25 audio tokens), 
which then requires much lower GFLOPS than padding of tens of seconds. 
We further show that hush words cannot be trivially constructed (e.g., as white noise)
but instead must be trained alongside the model.

Our second idea is to accelerate beam search by reusing the results of previous inference rounds as references.
While this shares the spirit of speculative execution for LLMs and speech~\cite{yang2023inference, kim2023speculative,wang2024turbochargespeechunderstandingpilot,huggingface_whisper_speculative}, 
SSP poses new challenges:
(1) the input audio buffer, a sliding window over the streaming audio, has its start/end periodically advanced (at an uneven pace) across consecutive processing rounds, resulting in \textit{misaligned} output sequences across rounds that cannot be trivially reused.
(2) such misalignment further causes \textit{timestamp tokens} to drift across rounds; 
this would disrupt the speculative algorithms. 
(3) special tokens (e.g., \_BEG\_) and punctuation 
further make existing speculative algorithms brittle. 
In response, our beam pruning builds atop autoregressive, speculative decoding and further incorporates novel designs: 
a search-based reference alignment between earlier decoding results and the current sequence; 
a selective token matching method to reduce the beam size;
and a fallback mechanism to mitigate errors from unreliable references.

Our third idea is to exploit the CPU/GPU \textit{pipeline parallelism}, 
as shown in Figure~\ref{fig:intro} (c): 
let the GPU execute encoding (high parallelism), 
and opportunistically offload decoding (low parallelism) to CPU; 
both processors form a \textit{pipeline}, working on different inputs in parallel. 
SSP makes designing an efficient CPU/GPU pipeline non-trivial: 
the stages of encoding and decoding exhibit a highly \textit{skewed} FLOPS ratio; 
the skewness varies across different models, input lengths, and hardware; 
there is no clear separation of resources: the decoding executed on GPU still requires nontrivial CPU time, competing with the CPU decoding in the meantime. 
Our new designs consist of: 
(1) a GPU-centric pipeline that offloads the decoding to CPU opportunistically.
(2) two separate, offline-tuned thread pools for encoding and decoding;
(3) offline profiling of models and hardware to determine the optimal threads allocation. 



\subsection{Results and contributions}
We implement a system called \sys{}, atop various client platforms. We test \sys{} on a long audio dataset, TED-LIUM3 long form version~\cite{hernandez2018ted} with data lengths spanning tens of minutes. 
On SSP task, \sys{} delivers comparable accuracy with per-word latency as low as 0.5 second with power consumption lower than 10 Watts. Compared to the popular foundation model SSP serving system, the on-device execution of \sys{} incurs 1.6$\times$--4.7$\times$ lower latency with little impact on energy consumption. 

This paper focuses on efficient inference of speech foundation models over streaming inputs, making them suitable for client devices. To this end, we contribute: 

\begin{myitemize}
\item Hush word,  
a novel method that cuts short the model input -- and therefore reduces the encoding computation -- without causing model hallucinations.

\item Beam pruning, a reference-based method to reduce the beam search computation with the help of the previous results, 
tailored to SSP scenarios with sliding audio windows.

\item CPU/GPU pipelining, a dynamic mapping of encoding and decoding between CPU and GPU. 
To our knowledge, we are the first to map SSP to heterogeneous processors. 


\end{myitemize}
	
We will make our code publicly available.


\section{Motivations}
\label{sec:motiv}

\subsection{Speech foundation models and SSP}

For the past several decades, speech understanding has been a key AI task, with research methodologies evolving from traditional signal processing~\cite{davis1952automatic} to Hidden Markov Models (HMM) ~\cite{Baker1975HMM} to modern deep learning approaches ~\cite{hori2017advances, peng2022branchformer, kim2024convolution}. 
Most of the existing SSP systems rely on
speech models that can operate on partial audio input and generate partial transcript output. 
However, lacking complete audio contexts and model generalizability, such systems often result in significant word errors. 

\textit{Foundation speech models}, as exemplified by OpenAI’s Whisper~\cite{radford2023robust}, are encoder-decode transformers. 
They are trained on massive datasets (e.g. 680K hours), 
more than 10$\times$ larger than 
those used for earlier models (often less than 50k hours).
The most notable benefit of foundation models is their generalizability: 
they deliver strong accuracy on unseen datasets (e.g. new vocabularies, accents, and challenging real-world conditions) without additional finetuning, 
which are contrasts with classic models. 
As such, they have the potential to offer a better trade-off between accuracy and latency, making SSP more practical for real-world deployment.



\subsection{Prior works and their limitations}
As foundation models present the state-of-the-art capability, 
there has been recent efforts on ``retrofitting'' them for SSP. 
\paragraph{The Whisper-Streaming system}
Notably, the Whisper-Streaming system \cite{machavcek2023turning} manages streaming audio by periodically updating an audio and transcript buffer, shown in ~\autoref{fig:intro} (b). During the operation, the new audio keeps extending the audio buffer, periodically with a pre-defined step length (if the step length is shorter than the model inference latency, the system will run with best effort, i.e., run the model inference back-to-back without waiting for longer audio segment ingested), the system runs Whisper model inference on the audio buffer and obtains current transcription result. The system compares the current results and the results in the transcript buffer. With LocalAgreement-2 \cite{Liu2020LowLatencySS}, the transcripts are confirmed when they appear in the consecutive model inference. The audio buffer is trimmed from the beginning when it exceeds a pre-defined threshold, and there are enough confirmed transcripts. 
The accuracy 
is decent -- comparable to the non-streaming Whisper system~\cite{radford2023robust}. 
In exchange of the accuracy is substantial computation overhead -- throughout the model execution. 
As a result, the system fails to deliver satisfying responsiveness on client devices.



\begin{table}[t]
    \centering
        \includegraphics[width=0.48\textwidth{}]{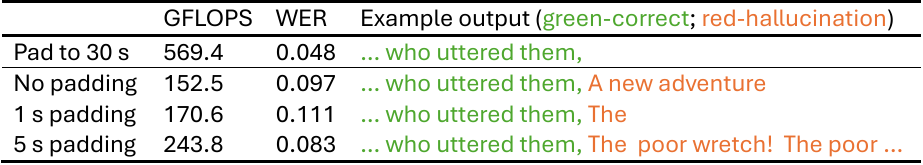}
        \caption{Speech Foundation Models typically expect the audio to be padded into a fixed length; shorter paddings tend to cause  hallucinations. GFLOPS is for encoding; WER (word error rate) is from Librispeech~\cite{panayotov2015librispeech} 
        dataset.}
        \label{tab:motivaudiohushtag}        
\end{table}


\paragraph{Inadequacy in encoding stage:} 
The audio buffer is fluctuating around the pre-defined length, e.g., 15 seconds. 
To accommodate the Whisper model requirement, the audio will be further padded into 30 seconds, as shown in ~\autoref{fig:intro} (b). 
The padding here causes substantial amount of redundant computation in encoding stage. 
As shown in ~\autoref{tab:motivaudiohushtag}, with default padding, the GFLOPS is 
3.7$\times$ more than the original input.
Attempting to feed the model with shorter inputs (i.e., 
with less padding) 
will result in significant hallucination and gibberish results, causing the accuracy to collapse~\cite{ggerganov2022streamdiscussion}.


\paragraph{Inadequacy in decoding stage:}
In SSP systems, there is significant redundancy in decoding stage. This occurs because the audio buffer, used as the model input, is updated each round but only with minor differences. This leads to substantial overlap in transcription results across rounds, creating redundancy in the decoding process. The beam search algorithm in decoding exacerbates this issue. It explores multiple hypotheses to find the best result—an effort often unnecessary in SSP.

Despite the redundancy, the overlap in the audio buffer is essential as it provides crucial contextual information for attention-based model
~\cite{vaswani2017attention}. Removing the overlap and processing audio segment-wise degrade accuracy, especially for shorter segments. A natural solution is to use previous results as references to guide the current beam search, similar to speculative decoding in LLMs. However, SSP introduces unique challenges. The results of consecutive inferences are misaligned due to the misalignment of audio buffers. During audio buffer updates, it is extended at the end with new audio and occasionally trimmed at the beginning. As a result, the transcript contents and the timestamp tokens are misaligned, compounded by inconsistencies in punctuation and special tokens.

\begin{table}[t]
    \centering
        \includegraphics[width=0.45\textwidth{}]{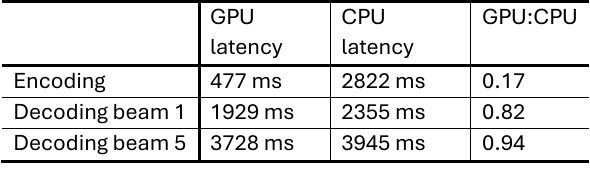}
        \caption{On M2 Pro chip, encoding and decoding show diverse characteristics and processor preferences. Model: Whisper medium. Decoding rounds: 150.}
        \label{tab:motivencdec}        
\end{table}


\subsection{SSP for CPU/GPU execution}
\label{sec:motiv:pipeline}


We are motivated by three key opportunities:

First, disparate levels of parallelism across processing stages. 
The encoding and decoding stages exhibit varying degrees of parallelism. The encoding stage is highly parallelized, involving a single pass to process 1,500 tokens. In contrast, the decoding stage operates more sequentially, consuming and generating several tokens per round (determined by the beam width). Despite the encoding stage processing an order of magnitude more tokens and requiring significantly higher GFLOPS, it often incurs longer delays than decoding when both are executed on GPU. This is reminiscent of spanning LLM prefill and decoding workloads across different processors~\cite{patel2024splitwise, wu2024loongserve, zhong2024distserve, xu2024empowering}.

Second, strong performance of modern CPUs on BLAS kernels. 
Modern CPUs, particularly those with specialized hardware accelerators, perform exceptionally well on BLAS kernels. A notable example is Apple silicon CPUs (later than M1), which include a built-in Apple Matrix accelerator ~\cite{corsix2022amx}. Our experiments in \autoref{tab:motivencdec} show that CPUs achieve comparable speeds to GPUs (only 1.1$\times$–1.2$\times$ slower) on \textit{decoding}, suggesting that GPUs could be reserved for \textit{encoding}, where they exhibit much stronger performance (up to $\sim$6$\times$ faster). However, on other platforms (e.g., Orange Pi, see evaluation), CPU performance can be significantly lower.

Third, pipelined execution over streaming input.
Unlike one-off LLM inference, streaming input provides a unique opportunity for pipelined execution. GPUs can process encoding for round \#$n$ while CPUs handle decoding for round \#$n-1$ simultaneously.



\section{Overview}
\label{sec:overview}


\subsection{System model}

We target client devices like mobile computers, smartphones, and embedded devices. 
The codebase of our system is self-contained; 
it can be installed on macOS, Android, Linux, and Windows without Python environment. 
Once deployed, the system runs entirely on-device without requiring the cloud. 

Our system primarily runs on mobile GPU; 
it can leverage idle CPUs, forming CPU/GPU pipelining for better performance.
In case GPU is unavailable, our system falls back to CPU-only execution, albeit with higher latency and smaller models. In our experience, today's mobile CPUs are not powerful enough to run SOTA speech models in real time. 


\subsection{Operations}

\begin{figure}[t!]
    \centering
	\includegraphics[width=0.45\textwidth]{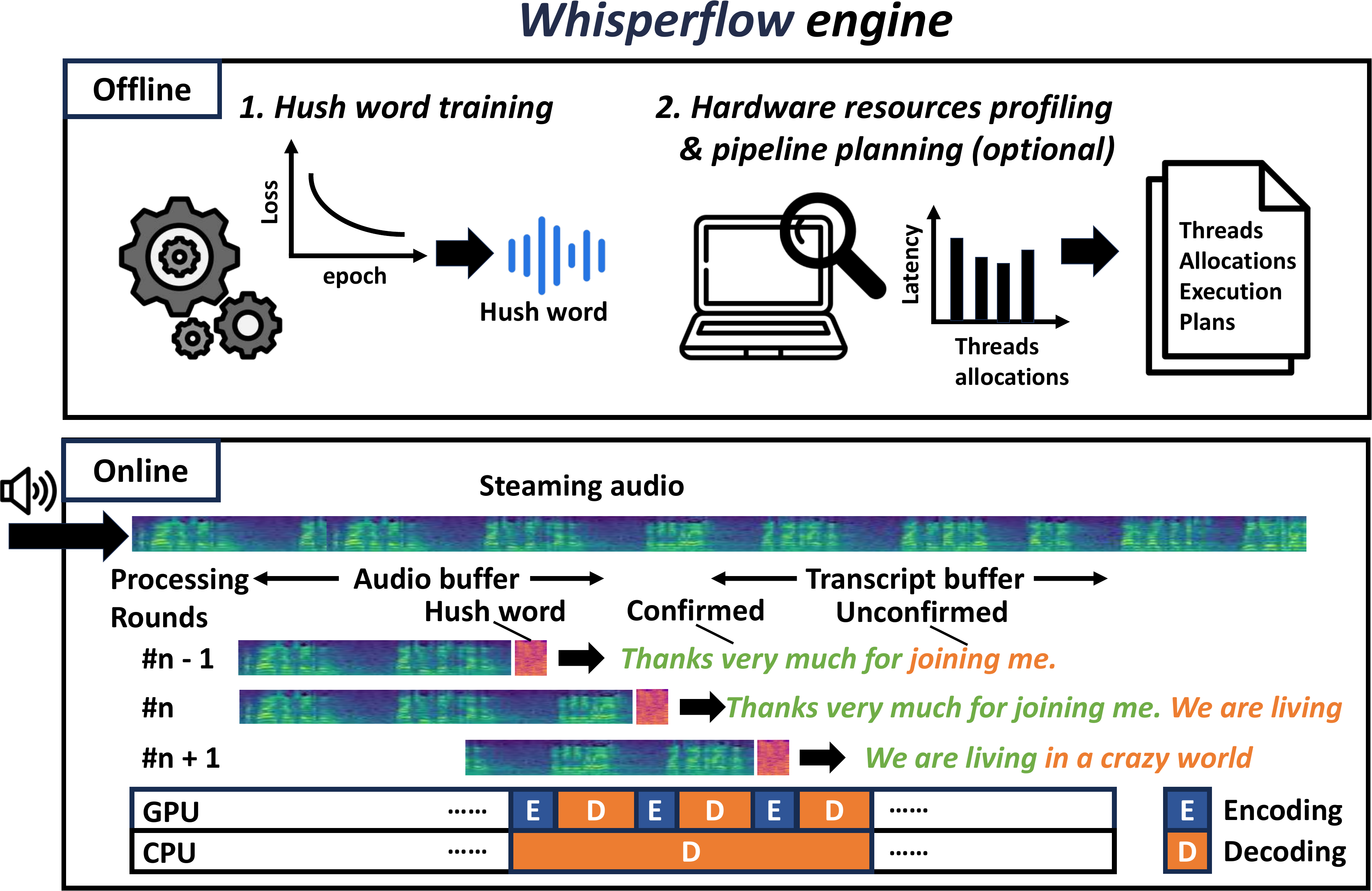}
	\caption{The \sys{} overview. 
    }
	\label{fig:pl}
	
\end{figure}


The architecture and operation of \sys{}, our proposed system, is shown in ~\autoref{fig:pl}. 

\paragraph{Offline development}
(1) We as system developers train special audio segments called hush word offline. 
We stress: only one hush word is needed for each foundation model (e.g., Whisper medium), 
as a one-time effort. 
(2) Optionally, system developers may profile the target platform hardware (e.g., Apple M2), 
which determines the optimal thread allocation for CPU/GPU pipelining; 
without such profiling, the system falls back to GPU-only inference.

\paragraph{Online operations}
(1) \textit{Processing rounds}. 
Our system continuously ingests streaming audio into the audio buffer. 
Periodically, the system takes a snapshot of the current \textit{audio buffer}, 
invokes the model for inference, and produces output tokens to its \textit{transcript buffer}. 
Such periodic inferences are referred to as ``processing rounds'' as shown in ~\autoref{fig:pl}. 
(2) \textit{Workload mapping}. During each round, the encoding, prompt prefilling, and token timestamping (using dynamic time warping, DTW~\cite{bellman1959mathematical}) are always mapped to GPU; opportunistically, the decoding is offloaded to CPU if it is idle.
(3) \textit{Transcript producing}. From the transcript buffer, the system materializes a segment of the transcript to users, only when the same segment is produced by two consecutive rounds (as exemplfied by green tokens in ~\autoref{fig:pl}). This algorithm, referred to as LocalAgreement-2~\cite{Liu2020LowLatencySS}, balances processing overhead and transcription accuracy.
(4) \textit{Audio buffer trimming}. To keep the audio buffer bounded, the system periodically trims the audio buffer by discarding samples that have already been materialized as transcript.


\noindent
\textbf{A processing round's latency} consists of the time for encoding an entire audio buffer, and for decoding the entire transcript sequence.
\textbf{A word's latency} consists of (1) the wait time (from the word's audio being ingested into the audio buffer to the system starts the next processing round); 
(2) the time of at least two rounds of processing, since the word must pass LocalAgreement-2 before its materialization. 

The goal of \sys{} is to minimize the average \textit{per-word latencies}. 
To do so, (1) it reduces per-round processing latency (encoding and decoding) 
and thus is able to transcribe the audio buffer with shorter intervals (i.e. shorter steps); 
(2) it minimizes accuracy degradation resultant from these optimizations. 
See \sect{impl} for exploration on step lengths. 

\subsection{Applicability}
Our implementation builds on SSP with the Whisper models. 
Our key ideas apply to encoder-decoder transformer speech models in general, which cover most of the SOTA speech models. In particular, we expect hush words to improve encoding efficiency for such models when transcribing short speeches in general scenarios. Beam pruning is more specifically bound to SSP workloads with sliding window audio buffers and beam search decoding. CPU/GPU pipelining also requires hardware with balanced CPU and GPU performance on specific stages of the workload.

\mbox{\sys{}} is designed for client devices, including but not limited to laptops, tablets, and mobile phones. 
For more powerful machines like servers, 
our techniques for computation reduction and CPU/GPU pipelining are still applicable. 
However, new challenges arise, e.g., hush words making batch processing less effective due to inconsistent input lengths.
For weaker devices like microcontrollers, running speech foundation models in real time remains an open challenge, and classic models may be more suitable.





\section{Design}
\label{sec:design1}
\subsection{Hush word}
\begin{figure}[t!]
	\centering
    \includegraphics[width=0.48\textwidth]{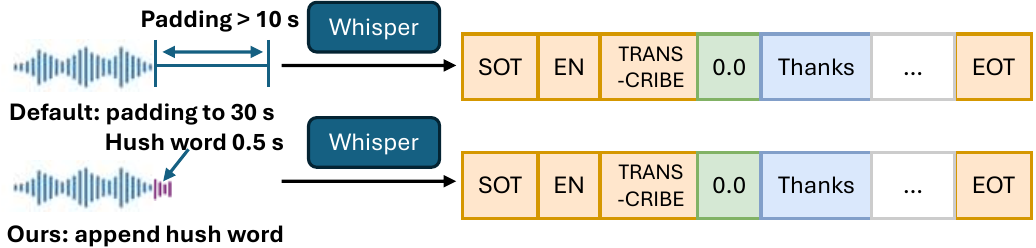}
	\caption{The hush word is a learnable audio segment designed to replace the long default padding that cause redundant computation.
	}
	\label{fig:designaudiohushtag}
	
\end{figure}



\subsubsection{Primer: Adversarial attack on foundation models}
As discussed in ~\autoref{sec:motiv}, the encoding stage in foundation models like Whisper is compute-intensive and handles fixed-length 30-second audio inputs. 
Shorter audio is padded to 30 seconds to avoid hallucination, as shown in ~\autoref{sec:motiv}. Naive padding methods (e.g., zeros or white noise) are ineffective.

Modifying Whisper models is impractical due to the high cost of finetuning. Full-parameter finetuning for a Whisper medium model ($\sim$750M parameters) demands over 30 GB VRAM and significant GPU hours. Parameter-efficient methods like LoRA reduce effort but may harm generalization ability, which is central to foundation models' appeal.

Adversarial attacks offer an alternative to altering model behavior without modifying the model. These attacks introduce adversarial segments into inputs to achieve specific outcomes. For instance, the "Muting Whisper" ~\cite{raina2024mutingwhisperuniversalacoustic} method prepends a 0.64-second adversarial segment to suppress transcription, training the model to output only the "end of transcript" token. This approach highlights the potential of adversarial attacks for behavior modification. However, our goal diverges from Muting Whisper, as we aim to spin the adversarial schemes to 
achieve beneficial objectives.

\subsubsection{New paradigm: hush word}
To address the problems mentioned above, we propose the ``hush word'', a novel model behavior-modifying scheme that helps foundation models eliminate redundant input padding.
As shown in ~\autoref{fig:designaudiohushtag}, we place our hush word -- a 
trained 
audio segment -- at the end of the original audio at the raw audio level with unique training targets. Several design choices need to be addressed.
\paragraph{Training scheme} 
The hush word audio segment is not manually designed but learned through training, which is conducted offline. Initially, the hush word is assigned a random value. During training, it serves as the only trainable parameter, appended to the audio data. The training process then performs forward and backward passes using the target foundation model, whose parameters remain frozen. The hush word as the learnable parameter is updated during training until it converges and becomes ready for use.
\paragraph{Training target}
Firstly, we set the training target to be the original transcript. This choice is straightforward, as we expect the model to produce the same transcript as the default output. In practice, the transcript is tokenized and formatted by prepending $\rm [StartOfTranscript], [language], [transcribe]$ and appending $\rm [EndOfTranscript]$. Consequently, the training target can be represented as $\rm y_{0} = [StartOfTranscript],$ $ \rm [language], [transcribe], [token\ 1-n], [EndOfTranscript]$.
\paragraph{Hush word position}
Secondly, we place the hush word at the end of the original audio, unlike Muting Whisper, which positions it at the beginning. This choice aligns with our intuition that the model should transcribe all audio content and be signaled by the hush word at the end to terminate transcription. Let the hush word have a length $T$, represented as ${\rm h} = h_{1:T}$. To hush a speech signal ${\rm x} = x_{1:N}$, the resulting combination can be expressed as $\rm x \oplus h$ 
to concatenate the audio and the hush word. 
The  training optimization objective 
\mbox{$\hat{h}$} (which is the final hush word) 
is defined as: 
\begin{center}
    \mbox{$\hat{h} = \mathop{\mathrm{argmax}}\limits_{\rm h} P\big(y_1 = {\rm y_0[1:]} \,\big|\, {\rm x \oplus h, y_0[:-1]}\big)$}
    
\end{center}

\paragraph{Hush word target level}
Thirdly, we choose to place the hush word at the raw audio level. There are two major target levels to consider: raw audio and log-mel spectrogram. Raw audio is a one-dimensional array, typically sampled at 16,000 Hz, where 1 second of audio corresponds to a vector of length 16,000. In contrast, a log-mel spectrogram is a two-dimensional representation computed from the raw audio, where 1 second of audio corresponds to a matrix with a shape of (100, 80).
We conducted an empirical study on hush word training at both target levels. For a 1-second equivalent hush word, both levels produced similar training loss. However, on the test set, the raw audio level achieved a lower word error rate. This is likely because the training results at the raw audio level provides a more consistent representation for the model compared to the log-mel spectrogram level, which may include irregular high or low frequency components that the model struggles to handle correctly.

\paragraph{Hush word length}
Lastly, we choose to keep the hush word length short, specifically 0.5 seconds. Empirically, during hush word training, a longer hush word can help achieve lower training and validation loss. However, during model inference, we find that a longer hush word does not necessarily result in a lower overall WER. By examining the transcripts generated with different hush word lengths, we observe that longer hush words introduce unique hallucinations, such as repeating certain capitalized phrases, particularly when the input audio is short. This behavior is reasonable, as appending a long hush word to short audio is more likely to confuse the model, even after training. Consequently, we select the 0.5-second hush word, which provides the best accuracy.

\paragraph{Offline training details}
With the design choices outlined above, we describe the general workflow for offline hush word training. The hush word is model-specific, meaning a dedicated hush word is trained for each model. Training is conducted on a small-scale audio dataset consisting of audio-transcript pairs.
We use a batch size of 1 during training, as we do not pad audio to different lengths but instead append a fixed-length hush word. For each data sample, the hush word is appended to the audio, and the corresponding transcript is prepared as described in the previous paragraph. Training continues until convergence, and we select a few of the best-performing epochs to compute the average final hush word.


\paragraph{Applicability and generalizability} 
The hush word technique applies broadly to speech foundation models with encoder-decoder Transformer architectures. It addresses the redundant padding challenge introduced by large batches and audio padding in training foundation models. Redundant padding stems from the large-scale pre-training process, where audio data of varying lengths must be padded to a uniform length to form large batches and improve training efficiency. While this training approach enhances generalizability, it also introduces significant padding redundancy that the hush word technique effectively mitigates. The proposed training scheme can be directly applied, though model-specific hush word tuning may be needed for best results.

\subsection{Beam pruning}


\subsubsection{Challenges}
Beam pruning leverages the inference result from the previous round as a reference to guide subsequent rounds, reducing the computational redundancy of beam search. In general, later rounds verify earlier results without performing beam search with the full beam size. Similar paradigms are observed in LLM speculative decoding and pilot inference in speech processing.

However, in the SSP system scenario, unique challenges arise due to the sliding-window-style audio buffer. As described in ~\autoref{sec:motiv}, the audio buffer grows with each round, appending new audio at the end while periodically cutting from the beginning as transcript segments are confirmed. This process creates misalignment at both the beginning and end of consecutive audio buffers, leading to several challenges that hinder speculative decoding and verification: (1) differences in transcript contents: The content of transcripts can differ between consecutive rounds at the beginning and the end, unlike speculative decoding case, which assumes identical expected contents. (2) timestamp variations: the misalignment of audio buffers result in inconsistent timestamp token results, complicating traditional verification schemes. (3) inconsistent punctuation and special token prediction: Whisper models often struggle to provide consistent punctuation and special tokens  predictions. This inconsistency poses additional challenges for verification.
\subsubsection{New design in beam pruning}
\begin{figure}[t!]
	\centering
    \includegraphics[width=0.48\textwidth]{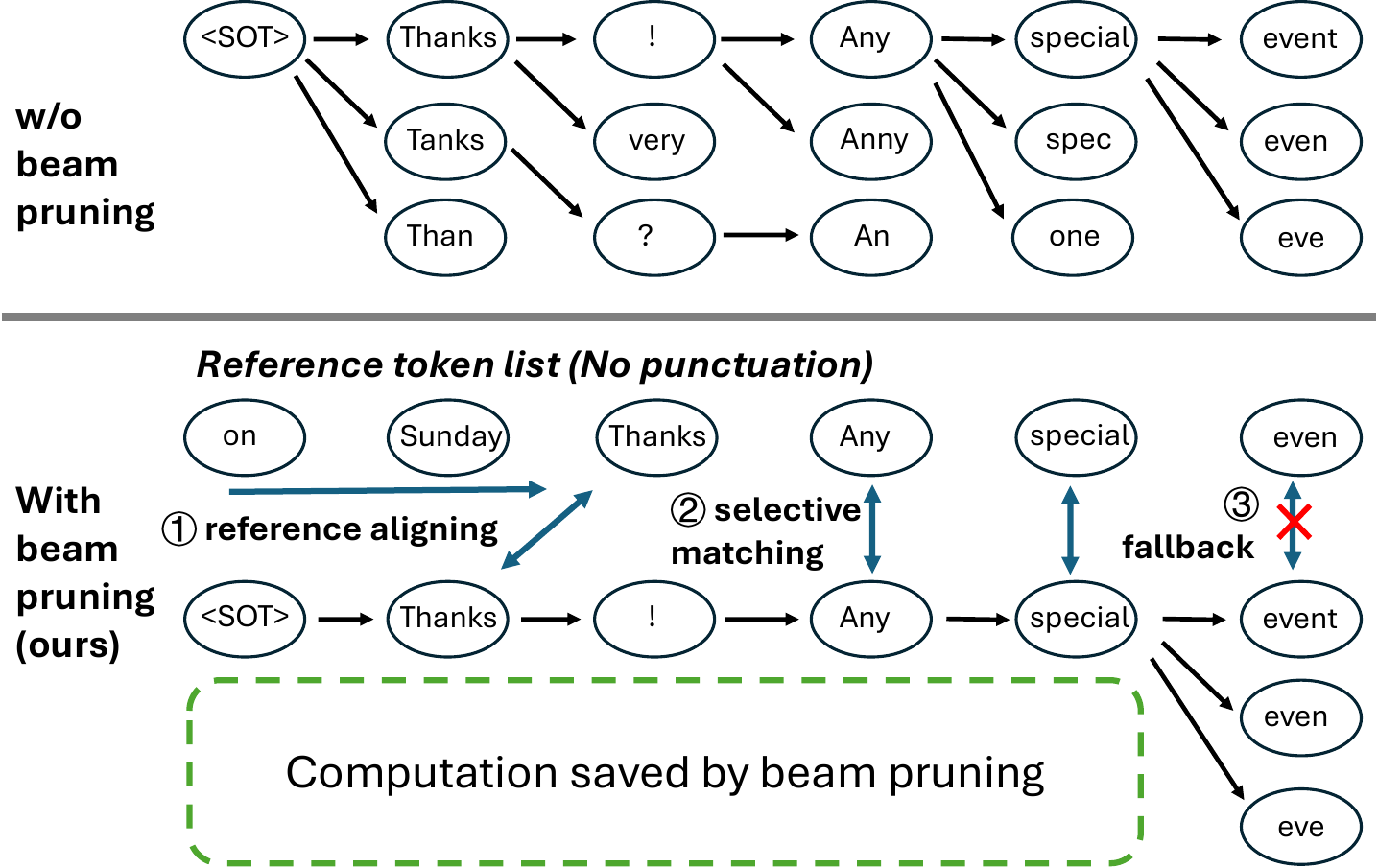}
	\caption{Our beam pruning reduces redundant beam search computation by aligning and matching the reference transcript tokens with a fallback mechanism.
	}
	\label{fig:desigbeampruning}
	
\end{figure}
To address the issues outlined above, we propose a novel beam pruning approach to minimize the redundancy in beam search decoding, even with partly misaligned references. This approach includes the following key components: (1) search-based reference alignment, (2) selective transcript token matching and beam reduction, and (3) a fallback mechanism
\paragraph{Search-based reference alignment}
At the beginning of beam pruning, we need to align the reference with the current decoding session. This alignment process occurs during the first round of decoding, where the beam size starts at 1. 
In the first round, we use the token with the highest probability to search through the reference token list 
to find the first match. Suppose the match occurs at token $i$; we then record this index for use in subsequent matching steps.
\paragraph{Selective transcript token matching}
Once we identify the starting point index $i$ in the reference token list, we proceed to selectively compare and match the highest-probability token from the current round with the reference tokens. If the best token matches the current reference token at index $i$, we maintain the beam size at 1, keep the best token as the new token, and update the reference index to $i+1$.

The matching process is selective. If the current best token is a punctuation, timestamp token, or any other special token, the matching for that round is skipped. In such cases, the beam size remains at 1, and the reference index is not updated. This ensures the potential misalignment caused by timestamps or punctuation does not interfere the matching. 
\paragraph{Fallback mechanism}
Beam pruning does not reduce the beam size indefinitely. Based on the results of reference alignment and transcript token matching, beam pruning will trigger a fallback to the original beam size under certain conditions. Two primary situations can trigger this fallback: (1) failure in reference alignment: if the alignment process at the start of decoding fails to find a match between the highest-probability token from the current decoding and the reference transcript tokens, the fallback mechanism is activated.
(2) mismatch during token matching: during the decoding session, if the current token is not a timestamp or punctuation token and fails to match the corresponding reference token, the fallback mechanism is triggered.
\paragraph{Applicability and generalizability} 
Beam pruning can be applied on systems with other models, as long as the system applies sliding window audio buffer which provides overlap in consecutive inference; and decoding with beam search algorithm, which is a common approach in speech tasks. Modern speech models with encoder-decoder architectures typically employ beam search to ensure high-quality results, which also introduces redundant computations aforementioned. Beam pruning effectively reduces these inefficiencies in the beam search process, making it well-suited for encoder-decoder architectures.

 
\subsection{CPU/GPU pipelining}

\subsubsection{Challenges}
As noted in ~\autoref{sec:motiv}, current SSP systems rely heavily on GPU-only execution, leaving CPU resources underutilized. However, leveraging CPU/GPU pipelining effectively presents several challenges: 
(1) diverse workload characteristics: the workloads in SSP include highly parallel tasks, such as encoding, prompt pre-filling, and DTW-based token timestamp estimation, as well as serial tasks, such as decoding with the beam search algorithm. Mapping these tasks onto GPU and CPU resources to form a bubble-free processing pipeline for SSP is a non-trivial problem. 
(2) resource contention: even when decoding runs on GPU, parts of the workload (e.g., sampling) are offloaded to CPU, requiring sufficient CPU resources. Poor thread allocation can lead to severe resource contention, further reducing efficiency. These unique challenges in SSP workloads necessitate a carefully tailored pipeline design.
\subsubsection{Pipeline for SSP}
\begin{figure}[t!]
	\centering
    \includegraphics[width=0.48\textwidth]{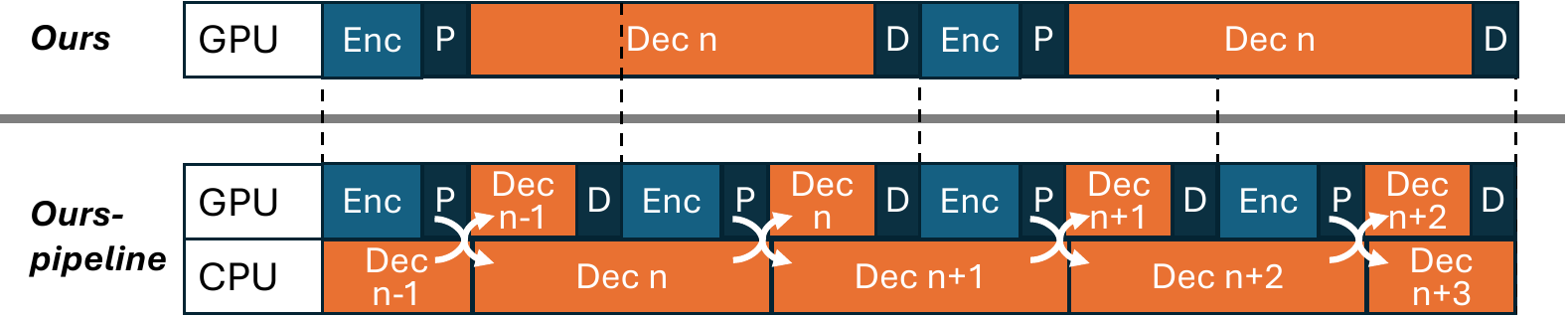}
	\caption{Our CPU/GPU pipelining opportunistically offloads decoding to (mostly) idle CPUs. Enc: encoding, P: prompt prefill, Dec: decoding, D: token timestamp estimation based on DTW ~\cite{bellman1959mathematical}. 
	}
	\label{fig:designpipeline}
	
\end{figure}
To address the identified challenges, we propose a new pipeline design optimized for SSP. 
\paragraph{Categorization of workloads}
SSP workloads involve consecutive model inference stages with token-level timestamp estimation. Each stage has unique computation characteristics and processor preferences: (1) encoding: 
for a typical 15-second audio buffer (including hush word), encoder processes $\sim$750 tokens. Encoding is highly parallel and strongly GPU-preferred. (2) prompt prefilling: processes $\sim$100 prompt tokens through the decoder. This is also a parallel workload with a strong GPU preference. (3) decoding: performs token-by-token decoding with beam search. For a 150-round average decoding session (5 hypotheses per round, reduced to 1 with beam pruning), this involves sequentially decoding $\sim$750 tokens. Decoding is highly sequential and less GPU-efficient. (4) DTW-based token timestamping: estimates token timestamps using the decoder cross-attention matrix for $\sim$150 decoded tokens. This workload is similar to prompt prefilling and benefits from GPU parallelism. Categorizing these workloads enables more effective resource allocation and reduces inefficiencies in the pipeline.
\paragraph{Opportunistically CPU offloads}
We let GPU executor process workloads on GPU, and CPU executors process workloads on CPU.
As mentioned above, we observe that encoding, prompt prefilling, and DTW token timestamp estimation are GPU-preferred, while decoding is less GPU-efficient. To address this imbalance, we propose a CPU offloading scheme that fully utilizes GPU while leveraging idle CPU resources.

As shown in ~\autoref{fig:designpipeline}, the pipeline operates as follows:
(1) workload distribution: we let the GPU executor processes encoding, prompt prefilling, and part of the decoding rounds. The CPU executor will handle the rest of the decoding rounds.
(2) pipeline coordination:
at th $n^{th}$ round, the GPU executor will start with $n^{th}$ encoding, prompt prefilling and then $n^{th}$ decoding, while the CPU executor do $n-1^{th}$ decoding. When the GPU executor finishes certain amount of the tokens (pre-defined before operation) of $n^{th}$ decoding, it will exchange the workload 
with the CPU executor. 
(3) pipeline advancement: once GPU completes $n-1^{th}$ decoding and DTW timestamp estimation, the pipeline advances to the $n+1^{th}$ round, and repeat the process mentioned above.
This design ensures maximal GPU utilization while limiting pipeline bubbles and idle time to the CPU, optimizing overall efficiency.
\paragraph{Threads allocations}
On typical client devices, we can allocate threads at a high level (e.g., GPU vs. CPU executors), but low-level CPU thread scheduling is managed by the operating system. Our experiment shows that GPU executor, especially during decoding, still require CPU resources for tasks like sampling. Allocating all CPU threads to the CPU executor causes resource contention and significant latency overhead (\autoref{sec:eval:designs}).

To address this, we propose an offline profiling method to optimize thread allocation: (1) thread pool setup: we divide all CPU cores (P cores and E cores) into two pools for GPU and CPU executors. (2) profiling: we use a short audio sample to test various thread allocation configurations and measure the per-word latency. (3) optimal configuration selection: base on the performance requirements and the profiling results, we select the final threads allocation configuration. By default we choose the one that offers the lowest per token latency. This offline profiling is a one-time effort conducted on devices before system deployment to ensure optimal resource utilization and minimal latency.

\paragraph{Applicability and generalizability}
The benefit of CPU/GPU pipelining relies on a basic assumption: on some of the workloads (usually decoding which is more serialized), the GPU and CPU have comparable performance. As a result, Apple silicons, along with other client platforms is generally suitable for the design. From the model perspective, the encoder-decoder architecture of modern speech models naturally divides processing into multiple stages with varying levels of parallelism. Some stages with serialized workloads can be effectively executed on idle CPUs, ensuring that available processing power is fully utilized. Building on this, the pipelining design distributes workloads according to their inherent parallelism, balancing tasks between the CPU and GPU. Thus, the pipelining design is well-suited for models with this architecture and effectively addresses their inefficiencies.
\paragraph{Limitations} 
Several challenges hinder the practical implementation of the pipelining design in real-world scenarios. In practice, the pipeline design cannot always keep the CPU fully occupied, as it must occasionally wait for the GPU. Additionally, on mobile SoCs, the computational power of CPU remains weak, which limits the CPU decoding efficiency. As a result, deploying the pipelining design remains a challenge.

\section{Key Implementation Details}
\label{sec:impl}
We implement our online system in 5K SLOC (C / C++) on top of the Whisper.cpp ~\cite{ggerganov2022whispercpp} 1.6.2 release. 
We implement our offline hush word training system in 300 SLOC (Python) on top of the Muting Whisper ~\cite{raina2024mutingwhisperuniversalacoustic} code base.

\paragraph{Hush word training}
We train the hush word for Whisper base (74M), small (244M), and medium (769M) models on the LibriSpeech dataset~\cite{panayotov2015librispeech}, with hush word lengths set to 0.5 second at the raw audio level 
(1-d tensor with shape (8000,))
. Training on an Nvidia A100 took 72 hours for the medium model, while training on an Nvidia A40 took 48 hours for the base and small models.

\paragraph{Offline profiling for the pipeline execution}
We profile the pipeline version of the system with different thread allocations. Suppose the CPU has n cores in total, the profiling will start from 5 cores for the CPU executor, and increase by 1 every time. The GPU executor uses the rest but one cores. 
We test all possible settings and choose the setting with lowest per-word latency for deployment. 

\paragraph{Operation details}
During system operation on the Apple devices, we set the system step length to 0.01 second to force the system execute with best effort to deliver the lowest latency. The long-form audio more than 5 minutes are chunked into 5-minute  segments. The system operates on these 
5-minute audio segments.

\noindent
\textbf{Step length}, 
the interval between two consecutive processing on the audio buffer, 
has the following key impacts. 

\begin{myitemize}
    \item \textit{Shorter steps tend to reduce per-word latency}.
    Long steps give abundant time for the system to process the audio buffer in each round, 
    although audio samples are buffered longer before processing, which inflates per-word latency. 
    As the step length reduces, 
    the audio wait time decreases, which reduces the per-word latency. 
    As the step length drops below the processing latency of a single round, 
    the system will run processing rounds back-to-back, 
    with each round consuming the most recent snapshot of audio buffer; 
    we refer to this mode as ``\textbf{best effort}''. 
    Note that it is difficult to determine the shortest possible step length offline:
    as the audio inputs and transcripts vary in length, the processing time per round also varies dynamically. 

    \item \textit{Shorter steps tend to degrade model accuracy}. 
    With fewer new audio samples added to the buffer in each round, 
    the inputs to two consecutive rounds will share longer spans with less new information from newly ingested audio, which in turn induces the model to produce more similar transcripts -- including erroneous words shared by these transcripts. 
    As such, the same erroneous words would recur in outputs from consecutive rounds, 
    pass the LocalAgreement-2 check, 
    and materialize 
    as the final transcript.
\end{myitemize}
\sect{eval} will evaluate different step lengths.


\section{Evaluation}
\label{sec:eval}

\begin{table}[t!]
    \centering
        \includegraphics[width=0.48\textwidth{}]{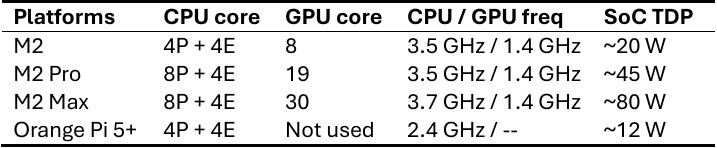}
        \caption{Arm platforms used in evaluation.         
        Orange Pi's GPU (Arm Mali) was not used in evaluation because of limited support on it. 
        TDP refers to thermal design power; our actual power is much lower (\autoref{sec:eval:e2e}) 
        }
        \label{tab:platformspecs}
\end{table}


\subsection{Methodology}

\paragraph{Test platforms} As shown in ~\autoref{tab:platformspecs}, we test \sys{} 
on a variety of Arm chips targeting client devices 
(Apple M2 series chips in 5nm, Orange Pi 5+ with Rockchip RK3588 in 7 nm).  
The 
Apple silicon devices have been widely used in client computing. 
To us, they also provide a deliberated engineered integration of CPU and GPU on the same chip, 
giving us a platform for considering real-world constraints of heterogeneous platforms.
We use the command line tool \texttt{powermetrics} which gathers the data from the hardware counters. 

\paragraph{Dataset} We conduct our experiments on the TED-LIUM 3 dataset~\cite{hernandez2018ted}, 
a collection of TED talks characterized by its 
diversity and challenging nature, including variations in speaking styles, accents, and speeds. We use 
its long-form version, which is also in Whisper's evaluation~\cite{radford2023robust}.

We evaluate hush word performance on   
LibriSpeech ~\cite{panayotov2015librispeech}, 
TED-LIUM 3 short-form and FLEURS~\cite{conneau2023fleurs} datasets, 
commonly used short-form audio benchmarks, 
with audio samples ranging in length from 5 to 25 seconds. 

\paragraph{Metrics}
We follow the common practice. 

\underline{Accuracy.} We report word error rate (WER) ~\cite{jurafsky2000speech}: the discrepancy between the system output and the ground truth transcript, defined as count of errors (substitutions, deletions, insertions) normalized by sentence length. This is widely used in prior works~\cite{radford2023robust, peng2022branchformer, Zhang2020transformertransducer}.

\underline{Latency.} We report the average per-word latency: the average of all the elapsed time from a user uttering the specific word to the system emitting the word in transcript ~\cite{machavcek2023turning, Zhang2020transformertransducer}.

\paragraph{Model choices}
We test with three models from the Whisper family: 
medium (769M), small (244M) and base (74M). We pick them because they are suitable for client device execution and provide good trade-offs between latency and accuracy. They are also as the focus in prior systems like Whisper.cpp \cite{ggerganov2022whispercpp}. 
We leave out tiny (39M) and large (1.55B) because the large variant incurs too long latency on client devices; the tiny variant delivers unsatisfactory accuracy. 
For baseline without foundation models, we adopt a streaming transformer model finetuned on TED-LIUM 2 dataset \mbox{~\cite{rousseau2014enhancing}} with 27.5M parameters, which is officially provided by ESPnet \mbox{~\cite{watanabe2018espnet}} maintainers for the state-of-the-art Streaming transformer \mbox{\cite{streamingtransformer2021tsunoo}} system.


\paragraph{Baselines} 
We compare the following designs. 



\begin{myitemize}


\item \baseline{}: 
This is our re-implementation of the original Whisper-Streaming design~\cite{machavcek2023turning}, 
a state-of-the-art SSP serving framework known for strong accuracy.
While the author's codebase is based on PyTorch, 
our re-implementation in C++, atop Whisper.cpp~\cite{ggerganov2022whispercpp}, 
is much more efficient: 
we avoid the overhead of PyTorch and its sub-optimal Metal backend (MPS)~\cite{mlx2023}. 
This ensures a fair comparison against our own system in terms of efficiency.
Our re-implementation consists of approximately 1200 lines of code.

\item \stransformer{}: 
The Streaming-transformer \mbox{~\cite{streamingtransformer2021tsunoo}} SSP system with the streaming speech model aforementioned. We adopt this baseline to represent the state-of-the-art SSP system without foundation models. The system is served with ESPnet toolkit, which is the most popular speech processing toolkit among researchers.

\item  \ourss{}: \sys{} with the encoding/decoding stages executed in serial, only on GPU.

\item \oursp{}: \sys{} with the encoding/decoding stages executes in parallel on both GPU and CPU.
\end{myitemize}

Note that we leave out Whisper.cpp's naive streaming implementation
(as introduced in ~\autoref{sec:intro}) from the experimental comparision here: 
as its accuracy is significantly worse than the aforementioned baselines, 
 we consider it less desirable for actual deployment.


\subsection{End-to-end results}
\label{sec:eval:e2e}





\begin{figure}[t!]
	\centering
	\includegraphics[width=0.48\textwidth]{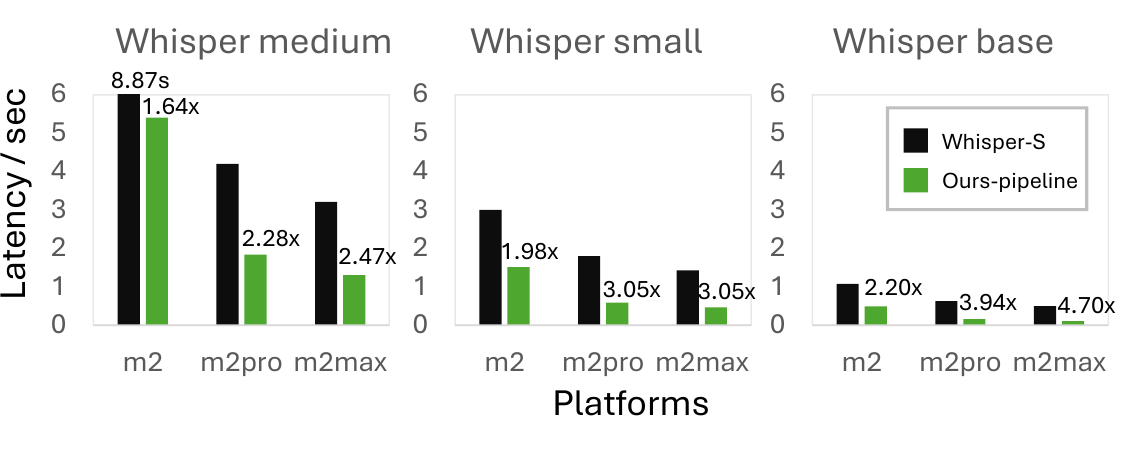}
	\caption{Compared to \textit{Whisper-S}, \oursp{} reduces per-word latency by 1.6$\times$--4.7$\times$, across three models and three devices. 
	Both systems run with best effort to achieve their lowest possible latencies, respectively. 
    }
	\label{fig:e2e}	
\end{figure}

\paragraph{Latency reduction}
As shown in ~\autoref{fig:e2e}, 
our system can serve Whisper models with low latencies across the devices in the experiments.
For smaller model (base), our per-word latencies can be as low as 0.2 second (on M2 Max); 
while for larger models with SOTA accuracy (medium), 
our system can keep the latency close to 1 second 
(on M2 Max, a popular chip choice for Macbook Pro). 
Such low latencies translate into good user experience: 
Prior studies ~\cite{poeppel2008speech, shangguan2021dissecting} show that sub-second latencies are close to or below user perception threshold, and latencies exceeding that would quickly degrade user experiences. 
For use cases where efficiency and latency are of higher priority, 
\sys{} can continuously execute Whisper base with only 1.2 seconds of latency 
on an entry-level Macbook Air, consuming only 7 Watts of power.
(details in the later ``\textbf{Power and energy}'' discussion).



Compared to the \baseline{}, our system can achieve lower latency. More specifically, on all the devices, with three different models, our system can 
reduce latencies by 1.6$\times$--4.7$\times$.
Our system can provide more pronounced saving on smaller models, and on more capable hardware.


\begin{figure}[t!]
	\centering
	\includegraphics[width=0.48\textwidth]{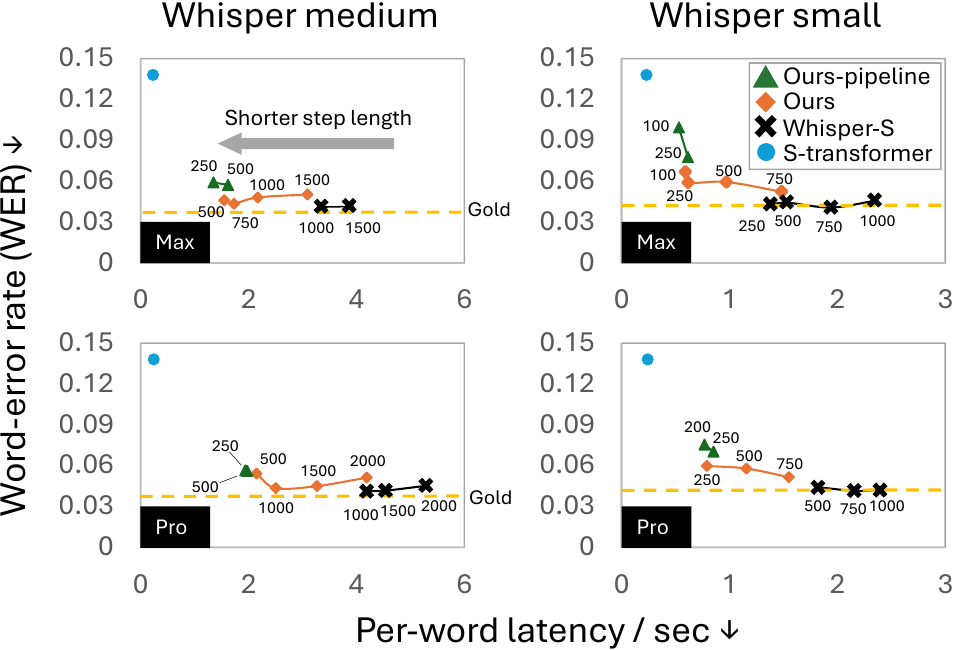}
	\caption{Compared to \textit{Whisper-S}, \textit{Ours} is capable of executing inferences at high rates (corresponds to shorter step length settings; numbers in ms, annotated);
	it hence delivers lower latency (x-axis) with minor accuracy degradation (y-axis). 
	\textit{Ours-pipeline} can further reduce the latency but may incur noticeable accuracy degradation for small models (right). Compared to \textit{S-Transformer}, \textit{Ours} significantly improves accuracy with the help of foundation models.
	Gold: accuracy of Whisper inference in non-streaming (batch) mode} 
	
	\label{fig:e2e2}
	
\end{figure}


Compared to S-Transformer, our system significantly improves accuracy (\mbox{~\autoref{fig:e2e2}}), reducing the word error rate from over 0.12 to below 0.06. While this comes with higher end-to-end latency due to the computational cost of speech foundation models, the trade-off is acceptable and makes the system practical for real-world use. This also highlights the superiority of foundation models.

On weaker devices, our system still shows significant latency reduction. 
Running Whisper tiny on all the CPU performance cores of Orange Pi 5+, 
our system can deliver 3.6 seconds per-word latency, 
2$\times$ lower than the 7.5 seconds latency of \baseline{} (not shown in figures).

\paragraph{Accuracy vs. latency}
Our systems incur minor accuracy drop compared to \baseline{}. 
As discussed in ~\autoref{sec:impl}, 
a key system parameter to control the accuracy/latency trade-off is the step length, i.e.,
time interval between two consecutive rounds of processing.
In general, shorter steps lead to lower latency (as the system produces output more frequently), but also to lower accuracy (as the model sees a shorter input context).
Figure~\ref{fig:e2e2} shows that as we reduce the step length for \baseline{}, 
the latency quickly reaches a lower bound, e.g., 3.3 seconds for Whisper medium on M2 Max; 
further reducing step lengths no longer reduces the latency.
The reason is: as the step length becomes close to, or shorter than, the system's per-round processing latency,
the system falls back to best-effort execution (see \S\ref{sec:impl} for description). 
By comparison, both \textit{ours} and \textit{ours-pipeline} deliver latencies much lower than what \baseline{} can achieve, 
thanks to our much lower per-round processing latency. 
Our WER is only marginally higher by 0.2\%-1.9\% on the medium model and 1.2\%-2.4\% on the small model, respectively, in most of the settings, regardless part of the accuracy degradation is the result of the shorter audio context; 
this WER is close to that of the \textit{non-streaming} Whisper inference with full context 
 (we consider as the ``gold'' accuracy). 
We also notice that with too short steps (e.g., less than 0.25 second), the model accuracy starts to degrade significantly due to lack of audio context. 




\begin{figure}[t!]
	\centering
	\includegraphics[width=0.48\textwidth]{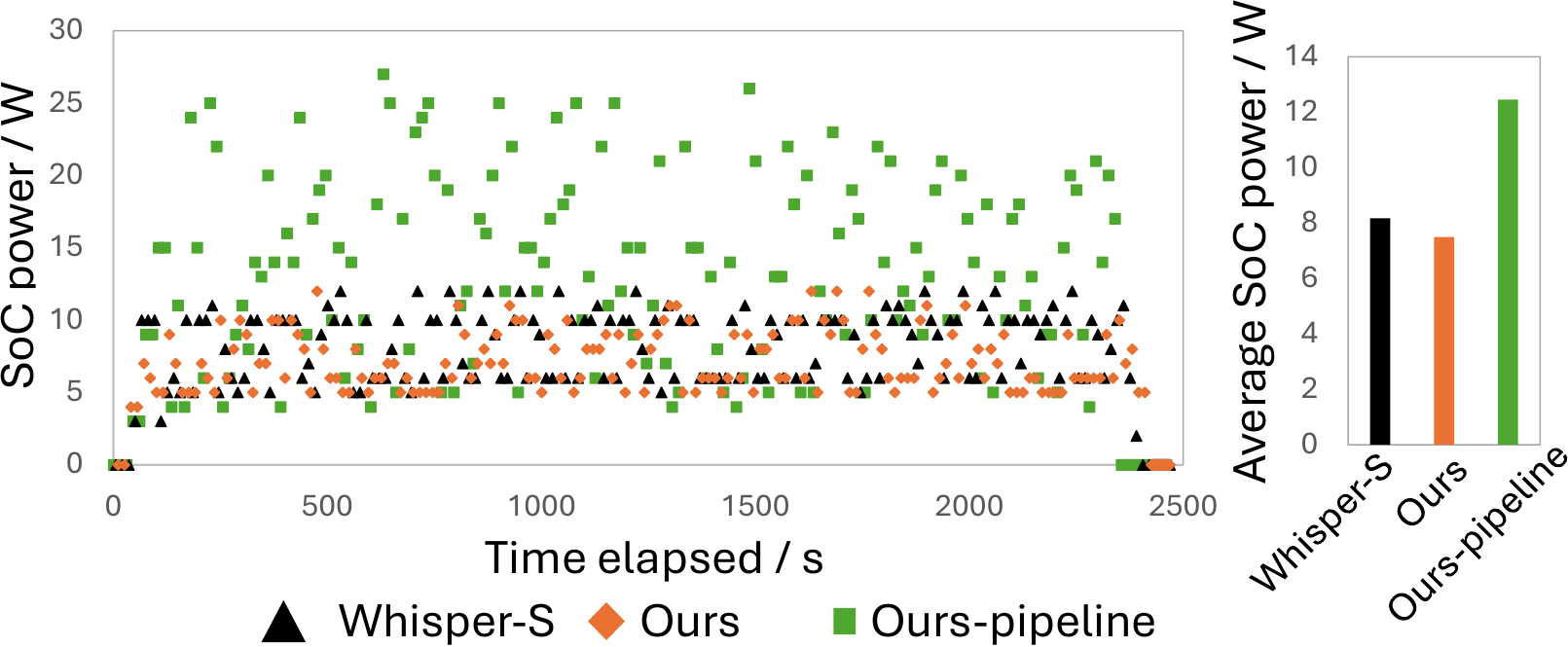}
	\caption{
		Power traces for processing 4-minute audio on M2 with Whisper small: 
		\ourss{} (using GPU only) reduces both system power and energy (besides per-word latency reduction), 
		while \textit{Ours-pipeline} (using CPU/GPU) may increase power and energy (albeit with more latency reduction). Measured by \texttt{powermetrics}.
		}
	\label{fig:energy}
	
\end{figure}

\paragraph{Power and energy}
\ourss{} reduces power. As the total time for processing a given audio input is almost the same between \baseline{} and \ourss{}, \ourss{} also reduces the energy consumption. 
This efficiency comes from our hush words and beam pruning, which reduce total GFLOPS. 
Figure~\ref{fig:energy} shows the SoC power for a 4-minute audio on the device with M2 chip and Whisper small model. 
The power of \ourss{} is $\sim$7 W, 8\% lower than \baseline{}  (note that \ourss{} also delivers 2$\times$ lower latency). 
This also means \ourss{} can run continuously for around 7 hours, on a typical 52.6 Wh laptop battery~\cite{apple2024macbookspecs}. 
Meanwhile, \textit{Ours-pipeline} incurs 1.5$\times$ higher power because it utilizes idle CPUs.
It is discussed further at the end of Section~\ref{sec:eval:designs}.


\subsection{Key designs}
\label{sec:eval:designs}
\paragraph{Ablation study}
\begin{figure}[t!]
	\centering
	\includegraphics[width=0.48\textwidth]{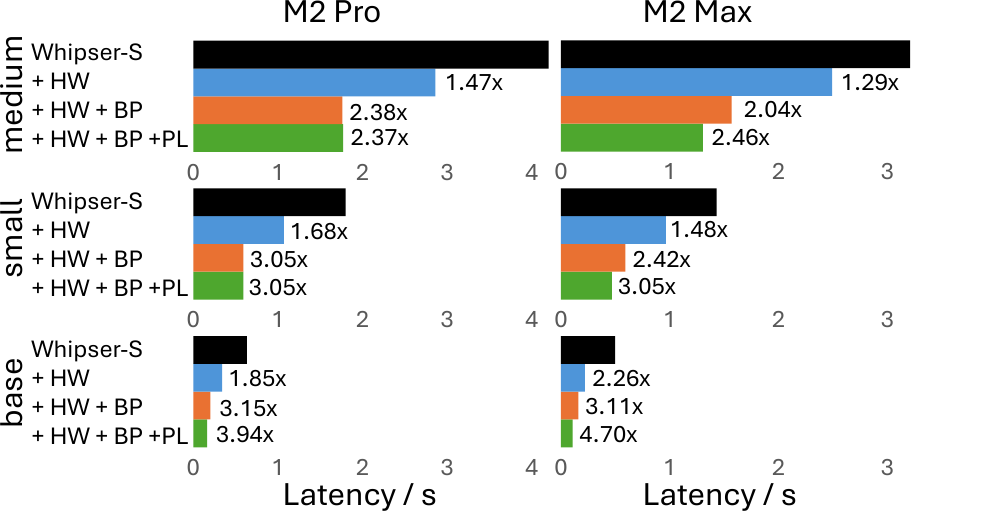}
	\caption{An ablation study of our system, showing that all its three optimizations contribute to lower latency comparing to the \baseline{}. It also shows the designs are effective on different model variants. HW: Hush word, BP: Beam pruning, PL: CPU/GPU pipelining 
    }
	\label{fig:speedupablation}
\end{figure}
As shown in ~\autoref{fig:speedupablation}, in most cases, all three designs contribute to overall latency reduction. (1) \textbf{hush word} reduces latency by 1.29$\times$--2.26$\times$. It brings more significant benefit on less capable devices, as encoding on these devices exhibits longer latency compared to decoding.
(2) \textbf{Beam pruning}, combined with \textbf{hush word}, reduces latency by 2.04$\times$-3.15$\times$. It provides greater latency reduction on more capable devices with larger models because the higher-quality references generated by these models are more effective for token matching in beam pruning. This is discussed further in the later paragraph. (3) \textbf{CPU/GPU pipelining} helps further accelerate the processing, reduces latency by 2.37$\times$-4.70$\times$ in total together with all other designs. On more capable devices with smaller models, the pipeline design offers greater benefits by better utilizing underutilized resources. 
However, using larger models on less capable devices introduces severe resource contention (mentioned in~\autoref{sec:design1}) in the pipeline, and increases latency.


\begin{figure}[t!]
	\centering
    \includegraphics[width=0.48\textwidth]{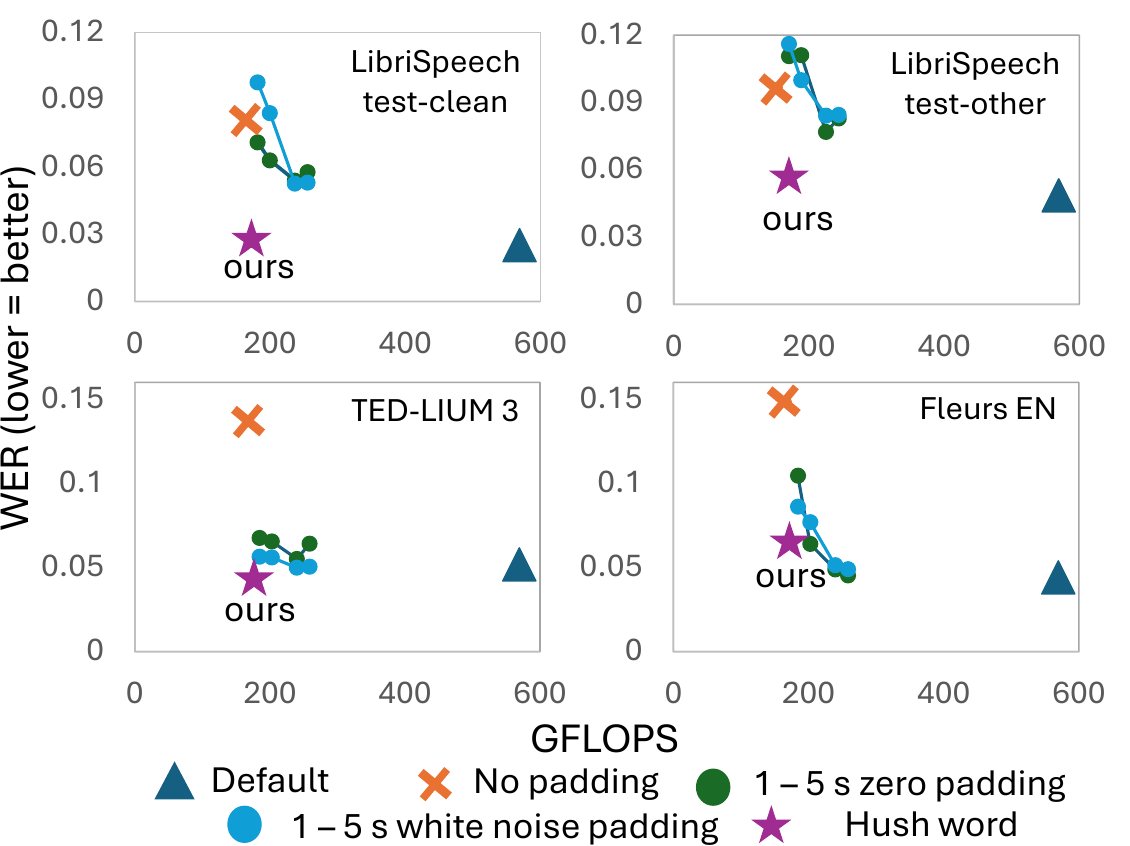}
	\caption{The hush word can effectively reduce the encoding GFLOPS while maintaining the accuracy, outperform other naive padding methods in most datasets.
	}
	\label{fig:audiohushtag}
	
\end{figure}

\paragraph{Hush word}
Our results show the efficacy and generalizability of hush word, across a variety of different datasets. 
As shown in ~\autoref{fig:audiohushtag}, the hush word is tested with non-streaming execution on LibriSpeech test-clean/other, TED-LIUM 3 short-form, and Fleurs datasets, compares with default execution (pad to 30 seconds), no padding, and 1 - 5 seconds zero/white noise padding. The hush word with 0.5 second length reduces the input length from 30 seconds to $\sim$10 seconds in these datasets. As a result, the encoding GFLOPS is reduced by 3$\times$ without major accuracy drop. The no padding setting shows lowest GFLOPS, but suffers from 5\%-10\% absolute accuracy drop (WER increase). Naive padding methods like padding with a certain length of silence or white noise audio may incur GFLOPS overhead while still deliver sub-optimal accuracy (lower WER). On LibriSpeech and TED-LIUM, comparing with 0.5 second hush word setting, with 5 seconds zero/white noise padding which incurs $\sim$40\% more GFLOPS, the accuracy is still 1\%--3\% lower. For Fleurs dataset, hush word still outperform 1 second zero/white noise padding, and shows similar accuracy with 2 seconds white noise padding, with 13.5\% less computation GFLOPS. Overall, for different datasets, different naive paddings may work and maintain the accuracy, but our hush word provides “Panacea” that can work in most of the cases without further profiling.

\paragraph{Beam pruning}
\begin{table}[t]
        \includegraphics[width=0.42\textwidth{}]{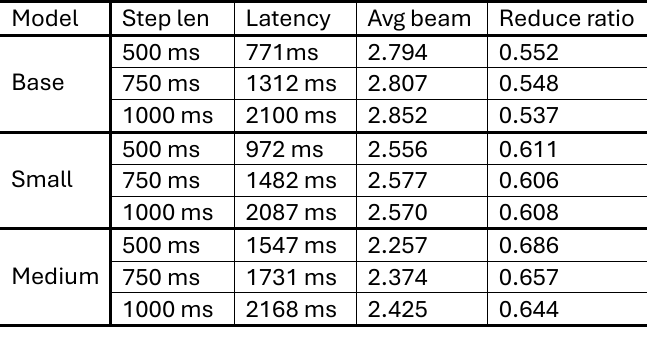}
        \caption{The beam pruning reduces the average beam size. The reduce ratio also shows that the beam size is reduced in the majority of the decoding rounds.}
        \centering
        \label{tab:beampruning}
\end{table}

The beam pruning design in our system shows capability of reducing the redundant computation in decoding process. Along with the ablation study, we check the beam pruning ratio of decoding rounds. As shown in ~\autoref{tab:beampruning}, beam pruning reduces the beam size in 53.8\%--68.6\% 
of the decoding rounds, making the average beam size become 2.26--2.85 (from the original beam size of 5). The larger models show higher beam reduce ratio and lower average beam size, because they have better capabilities and thus can provide higher quality reference results for beam pruning. 

\begin{figure}[t!]
	\centering
	\includegraphics[width=0.48\textwidth]{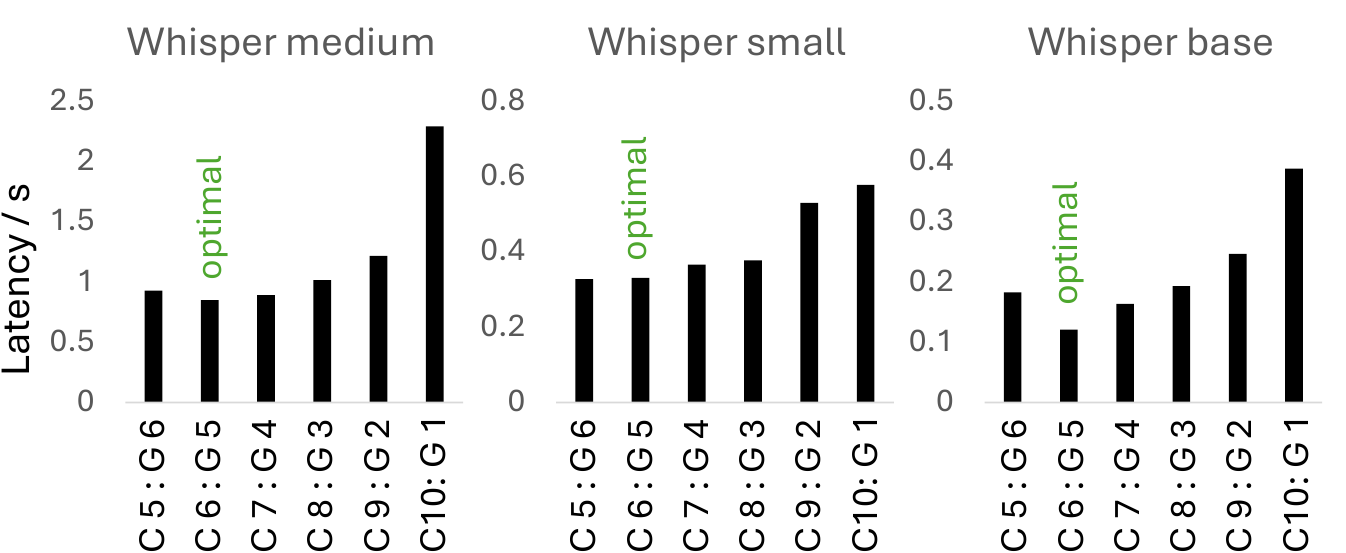}
	\caption{The CPU/GPU pipelining requires careful allocation of CPU threads between CPU and GPU executors. C\textit{x}:G\textit{y} means \textit{x} and \textit{y} CPU threads allocated to CPU and GPU executors, respectively. 
	Hardware: M2 Max; input for profiling: 4 minutes example audio. 
	}
	\label{fig:pipelinecontention}
	
\end{figure}

\paragraph{CPU/GPU pipelining}
\label{sec:eval:designs:pipeline}
reduces latency in most of the settings, as shown in ~\autoref{fig:e2e2} and ~\autoref{fig:speedupablation}; 
the latency reduction is more significant on more capable devices and smaller models. 
However, CPU/GPU pipelining also have clear downsides. 
(1) The latency reduction is sensitive to the CPU thread allocation; 
as shown in ~\autoref{fig:pipelinecontention}, 
the best allocation (C6:G5) has 40\% lower latency compares to bad allocations; 
the bad allocations may even underperform the GPU-only execution. 
The best choice of allocation depends on hardware architecture, model sizes, and complex thread contention. 
(2) Also shown in the results, the pipelining may degrade the accuracy noticeably, because with CPU/GPU pipelining, the system is able to achieve shorter step length, which results in shorter context. 
(3) The pipelining increases power and energy, whereas our GPU-only execution reduces power and energy (\autoref{fig:energy}). 

\subsection{Discussion}
\paragraph{Generalizability of the techniques}
Our experiments demonstrate that our system, incorporating all three techniques, generalizes well across different variants of Whisper models and various Apple Silicon devices. These techniques are also applicable to other models with similar encoder-decoder Transformer architectures, which are common in foundation models. In such models, hush word is particularly useful because padding during inference is an unavoidable consequence of the training scheme. Beam pruning enhances performance by efficiently managing the sliding window audio buffer, which is required for seamless integration of foundation models in streaming tasks. However, CPU/GPU pipelining imposes additional hardware performance requirements, which will be discussed later.
\paragraph{Lessons from CPU/GPU pipelining}
Despite our strong motivations \mbox{(\S\ref{sec:motiv:pipeline})} and our application of various pipelining tricks
, CPU/GPU pipelining is barely a ``free lunch'' as one may expect.
(1) The GPU executor also uses non-trivial \mbox{\textit{CPU}} cycles. It therefore may contend with CPU executors, as shown by our pipeline's sensitivity to thread allocations. 
(2) CPU is in general less energy-efficient than the GPU, in particular on ML workloads. 
(3) The addition of the CPU executor adds software complexity, which complicates code debugging and maintenance. 
On the brighter side, the pipeline could show higher benefit for the following factors: 
(1) Our pipeline, partitioned by encoder and decoder, is still coarse-grained; finer partitioning, e.g., at tensor levels, may see higher latency reduction; this, however, would further add to code complexity. 
(2) Future mobile CPUs will receive more cores and architectural features such as wider SIMD, narrowing their gap with the GPU of the same device. 
Overall, at the time of writing (Dec 2024), offloading to CPU for its \textit{larger memory} is a clearer choice \mbox{~\cite{ren2021zerooffload, ggerganov2022llamacpp, song2024powerinfer, huang2020swapadvisor, jung2023deepum}} ,
while offloading to CPU for its \mbox{\textit{idle cycles}} requires careful tradeoffs among latency, energy, and software complexity.

In real-world deployment, our system may run concurrently with other applications, causing fluctuations in hardware availability. One possible solution is to dynamically adjust the pipelining configuration based on runtime conditions. Offline profiling under dynamic hardware availability could help optimize configurations.
\paragraph{Target platforms}
As mentioned in \mbox{\autoref{sec:overview}}, our system targets client devices that can run foundation models with sufficiently low latency.
With our system, the least capable M2 chip achieves a usable latency of 1.5 seconds on Whisper small model, down from over 3 seconds without any optimization.
Less powerful devices experience even longer latencies, making foundation models impractical for real-world applications. As discussed in \mbox{\autoref{sec:eval}.2}, on Orange Pi 5+ with Whisper tiny, our system achieves a per-word latency of 3.6 seconds, outperforms 7.5 seconds with \mbox{\baseline{}}. However, the latency remains too high for real-world applications.
\section{Related work}
\label{sec:related}
\paragraph{Foundation models and streaming speech processing}
The emergence of the attention mechanism ~\cite{Bahdanau2014NeuralMT} inspires a generation of language models, from Transformer~\cite{vaswani2017attention} to more recent large-scale pre-training models like BERT ~\cite{devlin2019bert} and GPT ~\cite{radford2018improving, radford2019language, brown2020language}. The superior generalizability of these models is attributed to scaling law ~\cite{kaplan2020scaling}, which contributes to the success of GPT-3~\cite{brown2020language} and GPT-4~\cite{achiam2023gpt, ouyang2022training}. This paradigm has extended to speech processing, inspiring speech foundation models like Whisper ~\cite{radford2023robust} and OWSM ~\cite{peng2023reproducing, peng2024owsm, peng2024owsmctc}. They demonstrate great capability in speech-related tasks.
In this paper, we leverage the generalizability of these speech foundation models and propose a new system for streaming speech processing (SSP).
\paragraph{Streaming speech processing}
Streaming speech processing (SSP) is the task of processing and transcribing speech input in real time. For SSP tasks, approaches without foundation models~\cite{graves2012sequence,rao2017exploring,Zhang2020transformertransducer,streamingtransformer2021tsunoo} operate in a chunk-wise way with a short future audio context. As a result, they are limited by their reliance on partial data and constrained access to broader speech contexts. Trained on limited data, they also lacks the generalizability of SSP systems that leverage newer speech foundation models~\cite{machavcek2023turning,wang2024simul}.
We build our SSP system on top of Whisper-Streaming ~\cite{machavcek2023turning} where a series of novel techniques are proposed to improve the performance while maintaining its generalizability.

\paragraph{Adversarial attack on speech foundation models}
Adversarial attacks~\cite{szegedy2014intriguing} on speech models exploit vulnerabilities by introducing perturbations to input audio and manipulating the model behaviors. Earlier works ~\cite{carlini2018audio, alzantot2018did, Schnherr2018AdversarialAA} have succeeded in attacking end-to-end speech recognition models. With the development of the newer model architectures, universal attack ~\cite{neekhara2019universal} and attack that targeting commercial platforms~\cite{zheng2021black} are proposed. As the speech foundation models become prominent, they are the new target of adversarial attacks~\cite{olivier2022there,raina2024mutingwhisperuniversalacoustic,raina2024controlling}. Our work leverage the adversarial attack sample training scheme, and target at reducing the encoding redundancy in the speech foundation models.

\paragraph{LLM speculative decoding}
Speculative decoding (SD) to speech up LLM inference typically follows a ``generate first, verify later'' pattern~\cite{yang2023inference, xia2022lossless, sun2021instantaneous, kim2023speculative}. Inspired by similar techniques in LLMs, Whisper Speculative Decoding~\cite{huggingface_whisper_speculative} is proposed for speech processing. It utilizes a smaller, distilled model to generate a tentative result with low latency, and verifies it by a larger, fully-fledged model later. However, this SD method cannot be directly applied to reduce latency in our design because our system needs timestamp outputs for audio buffer management. Timestamp tokens may vary across consecutive inferences and disrupt our verification process.

\paragraph{Language models on heterogeneous processors}
Deploying language models to heterogeneous processors on client platforms has gained significant attention, which aims to utilize as much available resource as possible while meeting strict latency and power consumption requirements. It can be achieved via mapping transformer models between CPU and GPU~\cite{song2023powerinfer}, or transfer partial optimizer states to CPU during training~\cite{ren2021zerooffload}. There exist other works to address memory-bound scenarios, e.g., offloading a portion of the model parameters to CPU memory~\cite{ggerganov2022llamacpp, song2024powerinfer}, and storing key-value caches in CPU memory~\cite{huang2020swapadvisor, jung2023deepum}. In contrast, our work focuses on serving SSP tasks, and strives to balance between lowering latency and providing satisfying SSP performance.



\section{Conclusions}
We present \sys{}, a novel system for efficient inference of speech foundation models on streaming speech processing tasks. \sys{} contributes hush word, beam pruning, and a novel pipeline design tailored for client devices. With all optimizations combined, \sys{} reduces the per-word latency by up to 4.7$\times$ and enables operation with power consumption <10 W on portable devices.

\section*{ACKNOWLEDGMENT}
The authors were supported in part by NSF awards \#2128725, \#1919197, \#2106893, and Virginia’s Commonwealth Cyber Initiative. The authors thank the anonymous reviewers for their insightful feedback.


\bibliographystyle{plainurl}
\bibliography{bib/abr-short,bib/wrx}



\end{document}